# Metric-Independent Measures for Supersymmetric Extended Object Theories on Curved Backgrounds

Hitoshi NISHINO[1]) and Subhash RAJPOOT[2])

*Department of Physics & Astronomy*
*California State University*
*1250 Bellflower Boulevard*
*Long Beach, CA 90840, U.S.A.*

## Abstract

For Green-Schwarz superstring $\sigma$-model on curved backgrounds, we introduce a non-metric measure $\Phi \equiv \epsilon^{ij}\epsilon^{IJ}(\partial_i\varphi^I)(\partial_j\varphi^J)$ with two scalars $\varphi^I$ ($I = 1, 2$) used in 'Two Measure Theory' (TMT). As in the flat-background case, the string tension $T = (2\pi\alpha')^{-1}$ emerges as an integration constant for the $A_i$-field equation. This mechanism is further generalized to supermembrane theory, and to super $p$-brane theory, both on general curved backgrounds. This shows the universal applications of dynamical measure of TMT to general supersymmetric extended objects on general curved backgrounds.

---

[1]) E-Mail: h.nishino@csulb.edu
[2]) E-Mail: subhash.rajpoot@csulb.edu

## 1. Introduction

String theory, or more generally, theories of extended objects are believed to be the promising candidates for the unification of all interactions in nature [1]. For such theories, the desiderata is that there should be *no* fundamental scale involved in their Lagrangians. In other words, the real fundamental theory should involve *no* fundamental scale in its lagrangian, but instead it should arise at a later stage at the field-equation level, such as after spontaneous symmetry breaking.

One attempt to acquire such a system is found in the so-called 'New Measure' or 'Two-Measure Theory' (TMT). Conventional field theories in curved space time are typically described by actions with the measure $\int d^4x \sqrt{-g}$ which is metric-dependent, but otherwise invariant. However, it is possible to replace such a metric-dependent invariant measure by an alternative metric-*independent* measure, but is still invariant. Historically, such alternative-measure theories were first considered by Einstein and Rosen [2].

In 1990's, there were such formulations applied to superstring [3], and also to super $p$-branes [4][5]. Moreover, independent of these developments in 1990's, a new measure formulation in TMT was also given in [6][7] in terms of the scalar-density function $\Phi \equiv \epsilon^{\mu\nu\rho\sigma} \epsilon_{abcd} (\partial_\mu \varphi^a)(\partial_\nu \varphi^b)(\partial_\rho \varphi^c)(\partial_\sigma \varphi^d)$, where $\varphi^a$ are four scalars with the indices $a, b, \cdots = 1, 2, 3, 4$.

An important consequence of such an endeavor [6][7], when applied to the bosonic string [8], is the emergence of the string tension at the field equation level. In fact, the scalar-density function $\Phi$ yields the string-tension $T$. For the Polyakov-type bosonic string $\sigma$-model action [9]:

$$I_{\text{Bos}} = T \int d^2\sigma \left[ \tfrac{1}{2} \sqrt{-g}\, G_{mn}(X)\, g^{ij}(\sigma)\, (\partial_i X^m)(\partial_j X^n) \right] \ , \tag{1.1}$$

with the string tension $T \equiv (2\pi\alpha')^{-1}$, we consider the replacement of this lagrangian by

$$I_{\text{TMT-Bos}} = \int d^2\sigma \left[ \tfrac{1}{2} \Phi\, G_{mn}(X)\, g^{ij}(\sigma)\, (\partial_i X^m)(\partial_j X^n) \right] \ , \tag{1.2}$$

with the scalar-density function $\Phi \equiv \epsilon^{ij} \epsilon^{IJ} (\partial_i \varphi^I)(\partial_j \varphi^J)$ on two-dimensional (2D) worldsheet. The $G_{mn}(X)$ is the target-space metric, while $g_{ij}(\sigma)$ is the 2D metric.

However, the trouble with (1.2) is that the $g^{ij}$-field equation yields the unacceptable field equation $\Phi G_{mn}(\partial_i X^m)(\partial_j X^n) = 0$, leaving no dynamical freedom. This problem is solved by an additional term:

$$I_F \equiv \int d^2\sigma \left( \tfrac{1}{2} \epsilon^{ij} \chi F_{ij} \right) \equiv \int d^2\sigma \left( \tfrac{1}{2\sqrt{-g}} \epsilon^{ij} \Phi F_{ij} \right) \ , \tag{1.3}$$

where $\chi \equiv \Phi/\sqrt{-g}$, while $F_{ij} \equiv \partial_i A_j - \partial_j A_i$ is the field strength of the abelian gauge field $A_i$ on the 2D world-sheet. The effect of (1.3) is to provide a compensating term proportional to $g_{ij}\epsilon^{kl}\Phi F_{kl}$ for the $g^{ij}$-field equation, so the previous term $\Phi G_{mn}(\partial_i X^m)(\partial_j X^n)$ does *not* have to vanish by itself.

The most important conclusion of this bosonic-string formulation [8] is the determination of the string tension $T$ by the field equation of $A_i$ as[3])

$$\frac{\delta \mathcal{L}_{\text{Bos}}}{\delta A_i} = \epsilon^{ij}\partial_j \chi \doteq 0 \quad \Longrightarrow \quad \chi \equiv \frac{\Phi}{\sqrt{-g}} = \text{const.} \equiv T \quad \Longrightarrow \quad \Phi \doteq T\sqrt{-g} \ , \quad (1.4)$$

where $T$ is an integration constant interpreted as the string tension $T = (2\pi\alpha')^{-1}$. Despite the presence of the field $\Phi$ in (1.2), the original local Weyl symmetry of the action (1.1) is maintained in (1.2), because $\Phi$ transforms as a 'scalar-density' like $\sqrt{-g}$:

$$g_{ij} \to e^\Lambda g_{ij} \ , \quad g^{ij} \to e^{-\Lambda} g^{ij} \ , \quad \Phi \to e^\Lambda \Phi \quad \Longrightarrow \quad \Phi g^{ij} \to (e^\Lambda \Phi)(e^{-\Lambda} g^{ij}) = \Phi g^{ij} \ , \quad (1.5)$$

where $\Lambda = \Lambda(\sigma)$ is a local parameter. Note that the transformation rule for $\Phi$ is also consistent with the solution (1.4). Needless to say, the action $I_F$ is also invariant under the Weyl transformation (1.5), because of the special combination $(1/\sqrt{-g})\epsilon^{ij}\Phi$.

In ref. [10], this TMT mechanism [6][7] was further applied to superstring theory [1] in the Green-Schwarz (GS) superstring formulation [11] on the *flat* background. In our present paper, we consider the GS $\sigma$-model on *curved* 10D superspace background, including unidexterous fermions[4]) [12] with fermionic $\kappa$-symmetry [13]. Encouraged by the successful application to GS superstring, we further apply similar mechanism to supermembrane theory [14], and further to general super $p$-brane theories [4] on general curved backgrounds. The application of TMT formulation to *bosonic* $p$-brane theories was performed in [10], but *not* for *super* $p$-brane, the simplest case of which is supermembrane with $p = 2$. In our present paper, we carry out the TMT formulation for these *super* $p$-brane with general *curved* backgrounds.

This paper is organized as follows. In the next section, we present how the dynamical measure for TMT works for GS string $\sigma$-model. In section 3, we apply this mechanism to supermembrane theory. Section 4 is for the generalization to super $p$-branes. Concluding remarks is given in section 5.

## 2. GS Superstring $\sigma$-Model with Dynamical Measure

Before applying the new measure to GS superstring, we review the fermionic $\kappa$-invariance

---
[3]) We use the symbol $\doteq$ for a field equation or a solution, to be distinguished from algebraic equalities.
[4]) The meaning of 'unidexterous fermions' will be explained in the second paragraph in the next section.

[13] of the conventional Green-Schwarz superstring $\sigma$-model itself [11][12]. This procedure serves also as the preliminary notational arrangement.

The field content for the GS superstring $\sigma$-model on 10D superspace background [11][12] is $(V_{++}{}^i, V_{--}{}^i, Z^M, \psi_+{}^{(r)})$, where $(Z^M) = (X^m, \theta^\mu)$[5] is the 10D curved superspace background coordinates for GS string [11], while $(V_{(i)}{}^j) = (V_{++}{}^j, V_{--}{}^j)$ is the 2D zweibein. For 10D superspace curved coordinates, we use the indices $M = (m, \mu)$, where $m = 0, 1, \cdots, 9$ are for bosonic curved coordinates, while $\mu = 1, 2, \cdots, 16$ are for fermionic curved coordinates. For 10D superspace local coordinates, we use the indices $A = (a, \alpha)$, where $a = (0), (1), \cdots, (9)$ [6] are for local bosonic coordinates, while $\alpha = (1), (2), \cdots, (16)$ are for local fermionic coordinates. The index $(r) = (1), (2), \cdots, (32)$[7] on the unidexterous fermion $\psi_+{}^{(r)}$ is for the **32** of $SO(32)$ [12]. The word 'unidexterous' stands for the one-handed-ness of these fermions in 2D. Namely, all the 32 components of $\psi_+{}^{(r)}$ ($(r) = (1), (2), \cdots, (32)$) have the positive chirality, as its index $+$ indicates. On 2D world-sheet, the indices $i, j, \cdots = 0, 1$ are for the curved coordinates, while $(i), (j), \cdots = ++, --$ are for the light-cone local Lorentz coordinates. The necessity of these double indices is that the unidexterous fermion $\psi_+{}^{(r)}$ has the positive chiral index $+$ which is a single index, so that the bosonic coordinate $++$ (or $--$) is equivalent to the pair of two positive (or negative) chirality $+$ (or $-$). These facts have been well known as 2D features [1]. Note that the unidexterous fermion $\psi_+{}^{(r)}$ is in 2D, which is *not* directly related to the 10D-coordinates $Z^M = (X^m, \theta^\mu)$. Even though the range of the index $(r) = (1), (2), \cdots, (32)$ is twice as large as that of the fermionic-coordinate index $\alpha = (1), (2), \cdots, (16)$, the former is for the **32** of $SO(32)$ with *no* direct relationship with the latter for the fermionic coordinates $\theta^\mu$.

The action $I_{\rm GS}^{(0)} \equiv T \int d^2\sigma \, \mathcal{L}_{\rm GS}^{(0)}$ of GS superstring $\sigma$-model [11][12] has the string tension $T = (2\pi\alpha')^{-1}$ and the lagrangian

$$\mathcal{L}_{\rm GS}^{(0)} = + \tfrac{1}{2} V^{-1} g^{ij} \eta_{ab} \Pi_i{}^a \Pi_j{}^b - \tfrac{1}{2} \epsilon^{ij} \Pi_i{}^A \Pi_j{}^B B_{BA} + \tfrac{1}{2} V^{-1} \left( \psi_+{}^{(r)} \mathcal{D}_{--} \psi_+{}^{(r)} \right) \quad (2.1{\rm a})$$

$$= + V^{-1} \eta_{ab} \Pi_{++}{}^a \Pi_{--}{}^b - V^{-1} \Pi_{++}{}^A \Pi_{--}{}^B B_{BA} + \tfrac{1}{2} V^{-1} \left( \psi_+{}^{(r)} \mathcal{D}_{--} \psi_+{}^{(r)} \right) \ . \quad (2.1{\rm b})$$

The pull-back $\Pi_i{}^A$ is defined by $\Pi_i{}^A \equiv (\partial_i Z^M) E_M{}^A$. The $V$ is the determinant of the 2D zweibein $(V_{(i)}{}^j) = (V_{++}{}^a, V_{--}{}^b)$. The reason of negative power on $V$ in (2.1) is due to the definition $V \equiv \det(V_{(i)}{}^j)$, where the local index $(i)$ is used as the subscript [15]. The covariant derivative $\mathcal{D}_{--}$ is defined by

$$\mathcal{D}_{--} \psi_+{}^{(r)} \equiv V_{--}{}^i \partial_i \psi_+{}^{(r)} + \omega_i \psi_+{}^{(r)} + \Pi_{--}{}^A A_A{}^{(r)(s)} \psi_+{}^{(s)} \ . \quad (2.2)$$

---

[5] We are following the superspace notation in [15].

[6] The reason we use the parentheses is to distinguish them from local-coordinate indices.

[7] We need the parentheses for $(r), (s), \cdots$ to distinguish them from the local curved bosonic index $m, n, \cdots$.

The $\omega_i$ is the 2D Lorentz-connection, which drops out at the lagrangian level. The $A_A{}^{(r)(s)}$ is the Yang-Mills superfield in 10D whose $\theta = 0$ bosonic ($A = a$) component is $A_a{}^{(r)(s)}$, where the indices $(r)(s) = -(s)(r)$ are for the adjoint representation of $SO(32)$.

The action $I_{\text{GS}}^{(0)}$ is invariant under the fermionic $\kappa$-symmetry transformation [13][12][1]:

$$\delta_\kappa E^\alpha = -i(\sigma_a \kappa_{++})^\alpha \Pi_{--}{}^a \equiv -i(\sigma_{--}\kappa_{++})^\alpha \ , \qquad \delta_\kappa E^a = 0 \ , \tag{2.3a}$$

$$\delta_\kappa V_{++}{}^i = -2(\overline{\kappa}_{++}\Pi_{++})V_{--}{}^i + \tfrac{1}{2}\left(\overline{\kappa}_{++}\lambda^{(r)(s)}\right)\left(\psi_+{}^{(r)}\psi_+{}^{(s)}\right)V_{--}{}^i \ , \quad \delta_\kappa V_{--}{}^i = 0 \ , \tag{2.3b}$$

$$\delta_\kappa \psi_+{}^{(r)} = -(\delta_\kappa E^\alpha)A_\alpha{}^{(r)(s)}\psi_+{}^{(s)} \ , \qquad \delta_\kappa V^{-1} = 0 \ . \tag{2.3c}$$

Here $\delta_\kappa E^A \equiv (\delta_\kappa Z^M)E_M{}^A$, while $(\sigma^a)_{\alpha\beta}$ is the $\sigma$-matrix in 10D, and $(\sigma_{--})_{\alpha\beta} \equiv (\sigma_a)_{\alpha\beta}\Pi_{--}{}^a$. In (2.3b), we used the expression $(\overline{\kappa}_{++}\Pi_{++})$ for $\kappa_{++}{}^\alpha \Pi_{++\alpha}$ to save space. The $\lambda_\alpha{}^{(r)(s)} = -\lambda_\alpha{}^{(s)(r)}$ is for the gaugino in 10D in the adjoint **496** representation of $SO(32)$.

We give here the explicit total divergence form for $\delta_\kappa \mathcal{L}_{\text{GS}}$ that will be useful later:

$$\delta_\kappa \mathcal{L}_{\text{GS}}^{(0)} = -\nabla_{++}\left[V^{-1}(\delta_\kappa E^B)\Pi_{--}{}^A B_{AB}\right] + \nabla_{--}\left[V^{-1}(\delta_\kappa E^B)\Pi_{++}{}^A B_{AB}\right] \ , \tag{2.4}$$

leading to the invariance $\delta_\kappa I_{\text{GS}}^{(0)} = 0$.

As for the concept of 'general backgrounds', we add the following clarification. 'General backgrounds' imply that at least 10D space-time is curved by gravity with the non-trivial 10D metric $g_{mn}$. However, once gravity is introduced, for the consistency of the system with supersymmetry, all other supersymmetric partner superfields should be also introduced in a way consistent with $N = 1$ local supersymmetry in 10D. In other words, all 10D background superfields should be introduced consistently. They are *not* just limited to the NS-NS fields $g_{mn}$, $B_{mn}$ and $\varphi$. To be more specific, the $\theta = 0$ components corresponding to 10D component fields [16] are listed as $(e_a{}^m, \psi_a{}^\alpha, B_{ab}, \chi_\alpha, \varphi, A_a{}^{(r)(s)}, \lambda^{\alpha\,(r)(s)})$.

Once we have established (2.4) for the conventional Green-Schwarz $\sigma$-model [11][12], it is straightforward to confirm the $\kappa$-invariance of our new action with the new measure consisting of scalar fields $\varphi^I$ in place of the conventional measure from the metric.

To this end, we enlarge the field content to $(V_{++}{}^i, V_{--}{}^i, Z^M, \psi_+{}^{(r)}, \varphi^I, A_i)$. Here the new scalar field $\varphi^I$ has the index $I = 1, 2$, and $A_i$ is an Abelian vector field whose field strength is $F_{ij} \equiv \partial_i A_j - \partial_j A_i$. The scalar density function $\Phi$ is defined in terms of $\varphi^I$ by

$$\Phi \equiv \epsilon^{ij}\epsilon^{IJ}(\partial_i \varphi^I)(\partial_j \varphi^J) \ . \tag{2.5}$$

As is already known in the bosonic string case [8], a term linear in $F_{ij}$ is needed to cancel the unwanted term in the $V_{(i)}{}^j$-field equations. Moreover, this term is also needed from the viewpoint of $\kappa$-invariance of the total action, as will be seen next.

We propose our total action $I_{\text{GS}} \equiv \int d^2\sigma \, \mathcal{L}_{\text{GS}}$ to be

$$\mathcal{L}_{\text{GS}} = + \chi \mathcal{L}_{\text{GS}}^{(0)} + \frac{1}{2}\chi \, \epsilon^{ij} F_{ij}$$
$$= + \Phi \eta_{ab} \Pi_{++}{}^a \Pi_{--}{}^b - \Phi \Pi_{++}{}^A \Pi_{--}{}^B B_{BA} + \frac{1}{2}\Phi\left(\psi_+{}^{(r)} \mathcal{D}_{--}\psi_+{}^{(r)}\right) + \Phi F_{++,--} \ , \quad (2.6)$$

where $\chi \equiv V\Phi$. The $A_A{}^{(r)(s)}$ is the vector superfield for the YM-backgkound in 10D, while $(r)(s)$ are for the adjoint representation **496** of $SO(10)$, as in $\lambda_\alpha{}^{(r)(s)}$.

Our action $I_{\text{GS}}$ is invariant under the fermionic $\kappa$-transformation rule

$$\delta_\kappa E^\alpha = -i(\sigma_a \kappa_{++})^\alpha \Pi_{--}{}^a \equiv -i(\sigma_{--}\kappa_{++})^\alpha \ , \quad \delta_\kappa E^a = 0 \ , \quad (2.7\text{a})$$

$$\delta_\kappa V_{++}{}^i = -2(\overline{\kappa}_{++}\Pi_{++})V_{--}{}^i + \frac{1}{2}\left(\overline{\kappa}_{++}\lambda^{(r)(s)}\right)\left(\psi_+{}^{(r)}\psi_+{}^{(s)}\right)V_{--}{}^i \ , \quad (2.7\text{b})$$

$$\delta_\kappa V_{--}{}^i = 0 \ , \quad \delta_\kappa V^{-1} = 0 \ , \quad (2.7\text{c})$$

$$\delta_\kappa \psi_+{}^{(r)} = -(\delta_\kappa E^\alpha) A_\alpha{}^{(r)(s)} \psi_+{}^{(s)} \ . \quad (2.7\text{d})$$

$$\delta_\kappa A_i = -V_i{}^{--}(\delta_\kappa E^B)\Pi_{--}{}^A B_{AB} + V_i{}^{++}(\delta_\kappa E^B)\Pi_{++}{}^A B_{AB} \ , \quad (2.7\text{e})$$

$$\delta_\kappa \varphi^I = 0 \ , \quad \delta_\kappa \Phi = 0 \ . \quad (2.7\text{f})$$

The invariance $\delta_\kappa I_{\text{GS}} = 0$ is confirmed as follows. First, $\delta_\kappa \Phi = 0$ and $\delta_\kappa V = 0$ lead to $\delta_\kappa \chi = 0$, drastically simplifying the whole computation. This is because the variation $\delta_\kappa \mathcal{L}$ is only from $\delta_\kappa \mathcal{L}_{\text{GS}}^{(0)}$ and $\Phi \delta_\kappa F_{++,--}$. In particular, we already know the former as in (2.4). After a partial integration, the former yields a derivative on $\chi$, which is cancelled by the variation $\delta_\kappa F_{ij}$ again after a partial integration. Note that the invariance $\delta_\kappa I_{\text{GS}} = 0$ is *not* approximated one, such as only up to certain degrees in terms of $\psi_+{}^{(r)}$. In other words, our action $I_{\text{GS}}$ is confirmed to be $\kappa$-invariant to all orders. Thus we conclude that there is *no* problem for the $\kappa$-invariance of our action: $\delta_\kappa I_{\text{GS}} = 0$.

We next study all the field equations of $A_i$, $\psi_+{}^{(r)}$, $V_{++}{}^i$, $V_{--}{}^i$ and $\varphi^I$ in turn:

(i) $A_i$-Field Equation: This is the simplest one derived as

$$\frac{\delta \mathcal{L}_{\text{GS}}}{\delta A_i} = +\epsilon^{ij}\partial_j \chi \doteq 0 \quad \Longrightarrow \quad \partial_i \chi \equiv \partial_i(V\Phi) \doteq 0 \quad \Longrightarrow \quad V\Phi = \text{const.} \equiv T \ . \quad (2.8)$$

This implies that the combination $V\Phi$ is a constant $T$, i.e.,

$$\Phi \doteq T V^{-1} \ , \quad (2.9)$$

where the constant $T$ is interpreted as the string tension $T = (2\pi\alpha')^{-1}$.

(ii) $\psi_+{}^{(r)}$-Field Equation: The direct computation gives

$$\frac{\delta \mathcal{L}_{\text{GS}}}{\delta \psi_+{}^{(r)}} = +V^{-1}\chi(\mathcal{D}_{--}\psi_+{}^{(r)}) + \tfrac{1}{2} V^{-1}\psi_+{}^{(r)} V_{--}{}^i \partial_i \chi \doteq 0 \quad \Longrightarrow \quad \mathcal{D}_{--}\psi_+{}^{(r)} \doteq 0 \ . \qquad (2.10)$$

To get the last expression, have used (2.8).

(iii) $V_{++}{}^i$-Field Equation: The direct variation yields

$$\Pi_i{}^a \Pi_{--a} - V_i{}^{++} \Pi_{++}{}^A \Pi_{--}{}^B B_{BA} + F_{i,--} \doteq 0 \ . \qquad (2.11)$$

This equation yields, when multiplied by respectively $V_{--}{}^i$ and $V_{++}{}^i$,

$$\Pi_{--}{}^a \Pi_{--a} \doteq 0 \ , \qquad (2.12\text{a})$$

$$F_{++,--} \doteq -\Pi_{++}{}^a \Pi_{--a} + \Pi_{++}{}^A \Pi_{--}{}^B B_{BA} \ . \qquad (2.12\text{b})$$

The former is nothing but the conventional Virasoro condition, while the latter fixes the value of the new field strength $F_{++,--}$.[8] This situation is parallel to the bosonic case [8].

(iv) $V_{--}{}^i$-Field Equation: The direct variation yields

$$\Pi_i{}^a \Pi_{++a} - V_i{}^{--} \Pi_{++}{}^A \Pi_{--}{}^B B_{BA} + \tfrac{1}{2}\left(\psi_+{}^{(r)} \mathcal{D}_i \psi_+{}^{(r)}\right) - F_{i,++} \doteq 0 \ . \qquad (2.13)$$

When multiplied by $V_{--}{}^i$ and $V_{++}{}^i$, eq. (2.13) yields respectively

$$\Pi_{++}{}^a \Pi_{++a} + \tfrac{1}{2}\left(\psi_+{}^{(r)} \mathcal{D}_{++} \psi_+{}^{(r)}\right) \doteq 0 \ , \qquad (2.14\text{a})$$

$$F_{++,--} \doteq -\Pi_{++}{}^a \Pi_{--a} + \Pi_{++}{}^A \Pi_{--}{}^B B_{BA} \ . \qquad (2.14\text{b})$$

The former is nothing but the usual Virasoro condition with the unidexterous fermions, while the latter is consistent with (2.12b), as desired.

(v) $\varphi^I$-Field Equation: The direct computation gives

$$\frac{\delta \mathcal{L}_{\text{GS}}}{\delta \varphi^I} = +2\epsilon^{ij}\epsilon^{IJ}(\partial_i \varphi^J)\,\partial_j\Big[\,+\Pi_{++}{}^a \Pi_{--a} - \Pi_{++}{}^A \Pi_{--}{}^B B_{BA}$$
$$+ F_{++,--} + \tfrac{1}{2}\left(\psi_+{}^{(r)} \mathcal{D}_{--} \psi_+{}^{(r)}\right)\Big] \doteq 0 \ . \qquad (2.15)$$

This further yields

$$F_{++,--} \doteq -\Pi_{++}{}^a \Pi_{--a} + \Pi_{++}{}^A \Pi_{--}{}^B B_{BA} + M \doteq 0 \ , \qquad (2.16)$$

---

[8] The conventional Virasoro condition constrains only $\Pi_{++}{}^a \Pi_{++a}$ and $\Pi_{--}{}^a \Pi_{--a}$, but *not* $\Pi_{++}{}^a \Pi_{--a}$. The latter is *not* constrained in the conventional GS superstring [1].

due to the last term in (2.15) vanishing upon the $\psi$-field equation (2.10), while $M$ is a real integration constant. In our present TMT applied to GS superstring, or TMT applied to bosonic string [8], this constant $M$ is fixed to be *zero*, because of $V_{++}{}^i$ and $V_{--}{}^i$-field equations (2.12a) and (2.14a). This situation is different from more general TMT formulations [6][7], in which the constant $M$ remains to be non-zero in general.

To summarize, our system has the same field equations as the conventional GS superstring [11][12], together with new field equations. The examples of the former are (2.10), (2.12a) and (2.14a) [11][12]. Our new field equations are (2.8), (2.12b), (2.14b) and (2.16). The latter fixes the value of the new field strength $F_{++,--}$ and $M$, while the former results in the condition $\Phi \doteq T V^{-1}$, determining the string tension $T = (2\pi\alpha')^{-1}$. Both of these new field equations do *not* pose any new problem for GS string theory [11][12]. This situation is parallel to the aforementioned bosonic string [8] in the Polyakov-type formulation [9], and the GS superstring *flat*-background case [10].

We mention the fact that the equivalence between $I_{\text{GS}}^{(0)}$ for conventional GS [11][12] and our TMT generalization $I_{\text{GS}}$ is valid only at the classical level. Even for the conventional GS formulation [11][12], quantum computations are limited for general *curved backgrounds*, such as sigma-model $\beta$-function computations [17]. Since the quantum-level computations are highly non-trivial and need more arrangements for computations, it is beyond the scope of our present paper.

Even though TMT formulations for superstring were presented for *flat* background in [10], the importance here is that we have confirmed it also for GS superstring with general *curved* 10D superspace backgrounds [12].

## 3. Supermembrane with Dynamical Measure

As we have promised, we next apply this mechanism to supermembrane theory [14]. We first review the conventional supermembrane theory [14]. The field content of conventional supermembrane is $(Z^M, g_{ij})$, where $(Z^M) = (X^m, \theta^\mu)$ $(M = (m,\mu);\ m = 0, 1, \cdots, 10;\ \mu = 1, 2, \cdots, 32)$ are the 11D superspace coordinates, while $g_{ij}$ is the metric on the 3D world-volume [14].

The action $I_{\text{SM}}^{(0)} \equiv T \int d^3\sigma\, \mathcal{L}_{\text{SM}}^{(0)}$ of supermembrane has the lagrangian [14]

$$\mathcal{L}_{\text{SM}}^{(0)} = +\tfrac{1}{2}\sqrt{-g}\, g^{ij}\, \eta_{ab}\, \Pi_i{}^a \Pi_j{}^b - \tfrac{1}{2}\sqrt{-g} - \tfrac{1}{3}\epsilon^{ijk}\, \Pi_i{}^A \Pi_j{}^B \Pi_k{}^C B_{CBA}\ , \qquad (3.1)$$

where he $T$ is the membrane tension, while the $\Pi$'s represents the superspace pull-back $\Pi_i{}^A \equiv (\partial_i Z^M) E_M{}^A$ with the vielbein $E_M{}^A$ in the 11D superspace [18].[9]

---

[9] We use the notation in [15] in superspace.

The action $I_{\rm SM}^{(0)}$ is invariant under the fermionic $\kappa$-symmetry transformation rule [13]

$$\delta_\kappa E^\alpha = +\left[(I+\Gamma)\kappa\right]^\alpha \equiv (I+\Gamma)^{\alpha\beta}\kappa_\beta\ ,\quad \delta_\kappa E^a = 0\ ,\quad \delta_\kappa B_{ABC} = +(\delta_\kappa E^D)E_D B_{ABC}\ , \quad (3.2)$$

where $\delta_\kappa E^A \equiv (\delta_\kappa Z^M)E_M{}^A$, while $\Gamma$ is defined by

$$\Gamma \equiv \frac{i}{6\sqrt{-g}}\epsilon^{ijk}\Pi_i{}^a\Pi_j{}^b\Pi_k{}^c \equiv \frac{i}{6\sqrt{-g}}\epsilon^{ijk}\gamma_{ijk}\ , \quad (3.3)$$

with $\gamma_{ijk} \equiv \gamma_{abc}\Pi_i{}^a\Pi_j{}^b\Pi_k{}^c$. We also use symbols $\gamma_i \equiv \gamma_a\Pi_i{}^a$ and $\gamma_{ij} \equiv \gamma_{ab}\Pi_i{}^a\Pi_j{}^b$.

The explicit form of the variation $\delta_\kappa \mathcal{L}_{\rm SM}^{(0)}$ with surface term included will be useful for later purpose:

$$\delta_\kappa \mathcal{L}_{\rm SM}^{(0)} = +i\sqrt{-g}\left[\overline{\kappa}(I+\Gamma)\gamma^i\Pi_i\right] + \frac{1}{2}\epsilon^{ijk}\left[\overline{\kappa}(I+\Gamma)\gamma_{ij}\Pi_k\right]$$
$$- \nabla_i\left[\epsilon^{ijk}(\delta_\kappa E^C)\Pi_j{}^B\Pi_k{}^A B_{ABC}\right]\ , \quad (3.4)$$

where $\left[\overline{\kappa}(I+\Gamma)\gamma^i\Pi_i\right] \equiv \kappa^\alpha\left[(I+\Gamma)\gamma^i\right]_\alpha{}^\beta\Pi_{i\beta}$, etc. After using the relationships

$$i\gamma_i \doteq -\frac{1}{2\sqrt{-g}}\epsilon^{ijk}\gamma_{jk}\Gamma\ ,\quad \Gamma^2 \doteq +I\ , \quad (3.5)$$

we are left up only with the surface term in (3.4), confirming the invariance $\delta_\kappa I_{\rm SM}^{(0)} = 0$. Equalities in (3.5) are valid only up to the $g_{ij}$-field equation

$$g_{ij} \doteq \eta_{ab}\Pi_i{}^a\Pi_j{}^b\ , \quad (3.6)$$

also known as the 'embedding condition'. We also use the 11D superspace constraints [18]

$$T_{\alpha\beta}{}^c = +i(\gamma^c)_{\alpha\beta}\ ,\quad G_{\alpha\beta cd} = +\frac{1}{2}(\gamma_{cd})_{\alpha\beta}\ . \quad (3.7)$$

Our field content of TMT [6][7] applied to supermembrane [14] is $(Z^M, g_{ij}, \varphi^I, A_i{}^{IJ}, C_{ij})$. Here the scalar $\varphi^I$ ($I = 1, 2, 3$) is in the **3** of $SO(3)$ gauge group, similar to $\varphi^a$ used in TMT [6][7][10], while $C_{ij}$ is a tensor in 3D. Note that $A_i{}^{IJ}$ is the $SO(3)$ gauge field minimally coupled to $\varphi^I$. In other words, our system has the local $SO(3)$ symmetry with the $SO(3)$-covariant derivative $D_i\varphi^I \equiv \partial_i\varphi^I + A_i{}^{IJ}\varphi^J$. Compared with the GS superstring in section 2, the minimal coupling of the $SO(3)$ gauge field to $\varphi^I$ is new, whereas the Abelian gauge field $A_i$ in (2.6) is replaced by the tensor $C_{ij}$.

Our action $I_{\rm SM} \equiv \int d^3\sigma\,\mathcal{L}_{\rm SM}$ has the lagrangian

$$\mathcal{L}_{\rm SM} = +\frac{1}{2}\chi\sqrt{-g}\,g^{ij}\eta_{ab}\Pi_i{}^a\Pi_j{}^b - \frac{1}{2}\chi\sqrt{-g} - \frac{1}{3}\epsilon^{ijk}\chi\Pi_i{}^A\Pi_j{}^B\Pi_k{}^C B_{CBA} + \frac{1}{3}\epsilon^{ijk}\chi H_{ijk}\ , \quad (3.8)$$

with

$$\chi \equiv \frac{\Phi}{\sqrt{-g}} \ , \quad \Phi \equiv \epsilon^{ijk}\epsilon^{IJK}(D_i\varphi^I)(D_j\varphi^J)(D_k\varphi^K) \equiv \epsilon^{ijk}\epsilon^{IJK}P_i{}^I P_j{}^J P_k{}^K \ , \quad (3.9\text{a})$$

$$P_i{}^I \equiv D_i\varphi^I \ , \quad H_{ijk} \equiv \tfrac{1}{2}\partial_{[i}C_{jk]} \ . \quad (3.9\text{b})$$

Other than the presence of $SO(3)$-minimal couplings, this form is parallel to the scalar-density function used in TMT [6][7].

We next confirm the consistency of the field equations of our fields: $(C_{ij}, A_i{}^{IJ}, \varphi^I, g^{ij}, Z^M)$:

(i) The $C_{ij}$-Field Equation: The consequence of this simplest field equation is important:

$$\frac{\delta \mathcal{L}_{\text{SM}}}{\delta C_{ij}} = -\epsilon^{ijk}\partial_k \chi \doteq 0 \quad \Longrightarrow \quad \chi \equiv \frac{\Phi}{\sqrt{-g}} \doteq \text{const.} \equiv T \ . \quad (3.10)$$

This means that the membrane tension $T$ emerges as the integration constant for the $C_{ij}$-field equation, as one of our desired objectives.

(ii) The $A_i{}^{IJ}$-Field Equation:

$$\frac{\delta \mathcal{L}_{\text{SM}}}{\delta A_i{}^{IJ}} = +3\epsilon^{ijk}\epsilon^{[I|KL}\varphi^{|J]}P_j{}^K P_k{}^L$$

$$\times \Big[ +\tfrac{1}{2} g^{ij}\Pi_i{}^a\Pi_{ja} - \tfrac{1}{2} - \frac{1}{3\sqrt{-g}}\epsilon^{ijk}\Pi_i{}^A\Pi_j{}^B\Pi_k{}^C B_{CBA} + \frac{1}{3\sqrt{-g}}\epsilon^{ijk}H_{ijk} \Big] \doteq 0$$

$$\Longrightarrow \quad +\tfrac{1}{2}(\Pi_i{}^a)^2 - \tfrac{1}{2} - \frac{1}{3\sqrt{-g}}\epsilon^{ijk}\Pi_i{}^A\Pi_j{}^B\Pi_k{}^C B_{CBA} + \frac{1}{3\sqrt{-g}}\epsilon^{ijk}H_{ijk} \doteq 0 \ . \quad (3.11)$$

(iii) The $\varphi^I$-Field Equation:

$$\frac{\delta \mathcal{L}_{\text{SM}}}{\delta \varphi^I} = -3\epsilon^{ijk}\epsilon^{IJK}D_i\Big\{P_j{}^J P_k{}^K \Big[ +\tfrac{1}{2}(\Pi_l{}^a)^2 - \tfrac{1}{2} - \frac{1}{3\sqrt{-g}}\epsilon^{lmn}\Pi_l{}^A\Pi_m{}^B\Pi_n{}^C B_{CBA}$$

$$+ \frac{1}{3\sqrt{-g}}\epsilon^{lmn}H_{lmn} \Big] \Big\} \doteq 0 \quad (3.12)$$

In the usual TMT formulation [6][7], the covariant derivative $D_i$ is the ordinary derivative $\partial_i$, so that it commutes with $P_j{}^I P_k{}^K$. Eventually, the square bracket of (3.12) should be an arbitrary real constant $M$ [6][7]. However, the crucial difference here is that $D_i$ does *not* commute with the factor $P_j{}^I P_k{}^K$, so that the square bracket in (3.12) is *not* necessarily an arbitrary constant. Fortunately, the $A_i{}^{IJ}$-field equation (3.11) provides a stronger condition, such that the content of the square bracket in (3.12) vanishes. This is the advantage of the minimal coupling of the $SO(3)$-gauge field $A_i{}^I$ in our system.

(iv) The $g_{ij}$-Field Equation: This equation is the most crucial test, because we need the embedding condition $g_{ij} \doteq \Pi_i{}^a\Pi_{ja}$ [14]. In fact, we get

$$\frac{\delta \mathcal{L}_{\text{SM}}}{\delta g^{ij}} = +\tfrac{1}{2}\, \Phi\, \Pi_i{}^a\Pi_{ja} - \frac{1}{6\sqrt{-g}}\, \Phi\, g_{ij}\Big(\epsilon^{klm}\Pi_k{}^A\Pi_l{}^B\Pi_m{}^C B_{CBA} - \epsilon^{klm}H_{klm}\Big) \doteq 0 \ , \quad (3.13)$$

which is further simplified under (3.11) as

$$+ \frac{1}{2} \Pi_i{}^a \Pi_{ja} - \frac{1}{2} g_{ij} \left[ \frac{1}{2} (\Pi_k{}^a)^2 - \frac{1}{2} \right] \doteq 0 \ . \tag{3.14}$$

When the trace of this equation $(\Pi_i{}^a)^2 \doteq +3$ is again used in (3.14), it desirably produces exactly the embedding equation [14]

$$g_{ij} \doteq \Pi_i{}^a \Pi_{ja} \ . \tag{3.15}$$

(v) The $Z^M$-Field Equation: This field equation is eventually the same as in the supermembrane theory [14]:

$$T \nabla_i (\sqrt{-g}\, \Pi^i{}_a) + \frac{1}{2} T \epsilon^{ijk} (\gamma_{ab})_{\gamma\delta} \Pi_i{}^b \Pi_j{}^\gamma \Pi_k{}^\delta + \frac{1}{3} T \epsilon^{ijk} G_{abcd} \Pi_i{}^b \Pi_j{}^c \Pi_k{}^d \doteq 0 \ , \tag{3.16a}$$

$$i T \sqrt{-g} \left[ (I + \Gamma) \gamma^i \right]_{\alpha\beta} \Pi_i{}^\beta \doteq 0 \ . \tag{3.16b}$$

For reaching this final form, we have used the lemma (3.5), and the basic relationship

$$\delta \Pi_i{}^A = \nabla_i (\delta E^A) - \Pi_i{}^B (\delta E^D)(T_{DB}{}^A + \phi_{DB}{}^A) \ , \tag{3.17}$$

with the 11D Lorentz connection superfield $\phi_{DB}{}^A$ for an arbitrary variation $\delta E^A \equiv (\delta Z^M) E_M{}^A$. These field equations coincide with those in conventional supermembrane theory [14], and provide the supporting evidence of the consistency of our total system.

Note that our lagrangian (3.8) is reduced to the conventional supermembrane lagrangian (3.1) upon the use of $\chi \doteq T$ in (3.10). In particular, the $H$-linear term also disappears as a surface term, because under $\chi \doteq T$, it becomes a total divergence.

The explicit form of our fermionic $\kappa$-transformation rule is

$$\delta_\kappa E^\alpha = + [(I + \Gamma)\kappa] \equiv (I + \Gamma)^{\alpha\beta} \kappa_\beta \ , \quad \delta_\kappa E^a = 0 \ , \tag{3.18a}$$

$$\delta_\kappa B_{ABC} = + (\delta_\kappa E^D) E_D B_{ABC} \ , \quad \delta_\kappa C_{ij} = + (\delta_\kappa E^C) \Pi_i{}^B \Pi_j{}^A B_{ABC} \ , \tag{3.18b}$$

$$\delta_\kappa A_i{}^{IJ} = + \frac{1}{24 (\varphi^K)^2} \varphi^{[I} P_i{}^{J]} (\delta_\kappa g^{kl}) g_{kl} \ , \quad \delta_\kappa \varphi^I = 0 \ , \tag{3.18c}$$

while we do not specify $\delta_\kappa g_{ij}$ in our 1.5-order formalism, for the same reason already mentioned. Keeping this point in mind, and also using the result (3.4), we get the $\kappa$-invariance of our action

$$\delta_\kappa \mathcal{L}_{\text{SM}} \doteq - \epsilon^{ijk} \left[ \delta_\kappa C_{ij} - (\delta_\kappa E^C) \Pi_i{}^B \Pi_j{}^A B_{ABC} \right] \partial_k \chi$$

$$+ \left[ \delta_\kappa A_i{}^{IJ} - \frac{1}{24} \frac{1}{(\varphi^K)^2} (\delta_\kappa g^{kl}) g_{kl}\, \varphi^{[I} P_i{}^{J]} \right] \left( \frac{\delta \mathcal{L}}{\delta A_i{}^{IJ}} \right) = 0 \ , \tag{3.19}$$

where $(P^{-1})_J{}^i$ is the inverse matrix of $P_i{}^I$, satisfying $(P^{-1})_J{}^i P_i{}^I = +\delta_J{}^I$, and the first equality $\doteq$ in (3.19) symbolizes the usage of $g_{ij} \doteq \Pi_i{}^a \Pi_{ja}$ and a surface integration.

We have thus confirmed the invariance of our action $\delta_\kappa I_{\text{SM}} = 0$ under the fermionic $\kappa$-transformation (3.18) with general *curved* 11D backgrounds.

## 4. Generalization to Super $p$-Branes

Once we have understood the case of supermembrane, the generalization to *super $p$-branes* ($p \geq 3$) [4] is rather straightforward. For such a general $^\forall p$-brane formulation, the previous supermembrane for $p = 2$ becomes just the special case with the superspace [18] for 11D target space-time.

Our total action is $I_{p\text{B}} \equiv \int d^{p+1}\sigma \, \mathcal{L}_{p\text{B}}$, where

$$\mathcal{L}_{p\text{B}} = +\tfrac{1}{2}\chi\sqrt{-g}\, g^{ij}\,\eta_{ab}\,\Pi_i{}^a\Pi_j{}^b - \tfrac{1}{2}\chi\sqrt{-g}$$
$$- \frac{1}{p+1}\epsilon^{ijk}\chi\Pi_{i_1}{}^{A_1}\Pi_{i_2}{}^{A_2}\cdots\Pi_{i_{p+1}}{}^{A_d} B_{A_d\cdots A_2 A_1} + \frac{1}{p+1}\epsilon^{i_1 i_2\cdots i_d}\chi H_{i_1 i_2\cdots i_d} \quad , \tag{4.1}$$

with $d \equiv p + 1$ and

$$\chi \equiv \frac{\Phi}{\sqrt{-g}} \quad , \quad \Phi \equiv \epsilon^{i_1 i_2\cdots i_d}\epsilon^{I_1 I_2\cdots I_d} P_{i_1}{}^{I_1} P_{i_2}{}^{I_2} \cdots P_{i_d}{}^{I_d} \quad , \tag{4.2a}$$

$$P_i{}^I \equiv D_i\varphi^I \quad , \quad H_{i_1 i_2\cdots i_d} \equiv \frac{1}{d-1}\partial_{[i_1}C_{i_2\cdots i_d]} \quad . \tag{4.2b}$$

The fermionic $\kappa$-transformation rule is

$$\delta_\kappa E^\alpha = +[(I+\Gamma)\kappa] \equiv (I+\Gamma)^{\alpha\beta}\kappa_\beta \quad , \quad \delta_\kappa E^a = 0 \quad , \tag{4.3a}$$

$$\delta_\kappa B_{A_1 A_2\cdots A_d} = +(\delta_\kappa E^D) E_D B_{A_1\cdots A_d} \tag{4.3b}$$

$$\delta_\kappa C_{i_1 i_2\cdots i_d} = +(\delta_\kappa E^C)\Pi_{i_1}{}^{A_1}\cdots\Pi_{i_d}{}^{A_d} B_{A_d\cdots A_1 C} \quad , \tag{4.3c}$$

$$\delta_\kappa A_i{}^{IJ} = +\frac{1}{24(\varphi^K)^2}\varphi^{[I} P_i{}^{J]}(\delta_\kappa g^{kl})\,g_{kl} \quad , \quad \delta_\kappa\varphi^I = 0 \quad . \tag{4.3d}$$

As is easily seen, the previous supermembrane case is the special case of $p = 2$.

Even though TMT formulation was presented in [10] for super $p$-branes [4] for *flat* backgrounds, our present result is valid for general *curved* backgrounds in the target space-time.

## 5. Concluding Remarks

In this paper, we have applied TMT [6][7] to GS superstring [1] on general curved backgrounds, carrying out the objective to generate the superstring tension $T$ only as an integration constant, while it is absent from the fundamental lagrangian. This mechanism is further applied to supermembrane [14], and super $p$-brane theories [4], both on general curved backgrounds. The lagrangian of GS superstring is (2.6) with $\kappa$-invariance (2.7), that of supermembrane is (3.8) with $\kappa$-invariance (3.18), and that of super $p$-brane is (4.1) with $\kappa$-invariance (4.3).

The new feature of our result compared with [10] is that $\kappa$-invariances with TMT dynamical measures have been confirmed for supersymmetric extended objects, such as supermembrane, and more general super $p$-branes on general *curved* backgrounds. Even for GS superstring, we have added unidexterous fermions which were not treated in [10]. Even though the extra factor $\chi$ is multiplied by the conventional super $p$-brane lagrangian [14], all new contributions are cancelled by $\delta_\kappa C_{ij}$ and $\delta_\kappa A_i{}^{IJ}$.

In principle, we can apply TMT formulations [6][7] to D$p$-brane theory [19] in a similar fashion. In such a case, we need world-sheet Born-Infeld vectors. In practice, however, the required computation will be more involved beyond the scope of this Letter. We leave such formulations for future projects.

Our present results show that the dynamical measure in TMT [6][7] has general universal features applicable to supersymmetric extended objects, such as GS superstring [1], supermembrane [14], and super $p$-branes [4], on *general curved backgrounds*. Even though the generalizations to supersymmetric extended objects on general curved backgrounds seem straightforward, we have to confirm this conjecture by explicit computations. Based on our encouraging results, it is natural to expect that the basic properties of TMT dynamical measure [6][7] are universally applicable to even other (supersymmetric) extended objects.

## Acknowledgement

We are grateful to E. Guendelman for various important discussions. This work is supported in part by Department of Energy grant # DE-FG02-10ER41693.